\documentstyle[epsfig]{europhys}

\def\And{{\rm and\ }}

\def\stars{\bigskip\centerline{***}\medskip}

\newif\ifboo \boofalse

\def\Name#1{{\sc #1},}
\def\Vol#1{\ifboo Vol. {\bf #1}\else{\bf #1}\fi}
\def\Year#1{\ifboo #1\else(#1)\fi}

\def\Page#1{\ifboo {\rm p. #1}\else{\rm #1}\fi}

\begin{document}
\def\be{\begin{equation}}
\def\ee{\end{equation}}
\euro{0}{0}{0-0}{2002}
\Date{November 27 2001}
\shorttitle{Effects of an external drive ...}
\title{Effects of an external drive on the fluctuation-dissipation relation 
of phase-ordering systems}
\author{F.Corberi$^{\dag ,}$ \inst{1}, G. Gonnella$^{* ,}$ \inst{2}, 
E.Lippiello$^{\ddag ,}$ \inst{1} \And M.Zannetti$^{\S ,}$ \inst{1} }
\institute{
     \inst{1} Istituto Nazionale per la Fisica della Materia,
Unit\`a di Salerno {\rm and} Dipartimento di Fisica, Universit\`a di Salerno,
84081 Baronissi (Salerno), Italy \\
     \inst{2} Istituto Nazionale per la Fisica della Materia {\rm and}
Dipartimento di Fisica, Universit\`a di Bari 
{\rm and} Istituto Nazionale di Fisica Nucleare, 
Sezione di Bari, \\ 
via Amendola 173, 70126 Bari, Italy }
\rec{}{}
\pacs{
\Pacs{64}{75}{+g}
\Pacs{05}{70}{Ln}
\Pacs{83}{50}{Ax}
      }
\maketitle
\begin{abstract}
The relation between the autocorrelation $C(t,t_w)$ and 
the integrated linear response function $\chi(t,t_w)$ is studied
in the context of the large-$N$ model for phase-ordering systems
subjected to a shear flow. In the high temperature phase
$T>T_c$ a non-equilibrium stationary state is entered
which is characterized by a non-trivial fluctuation-dissipation
relation $\chi (t-t_w)=\tilde \chi(C(t-t_w))$. For quenches below $T_c$ 
the splitting of the order parameter field 
into two statistically independent components,  
responsible for the stationary $C^{st}(t-t_w)$ and aging 
$C^{ag}(t/t_w)$ part of the autocorrelation 
function, can be explicitly exhibited
in close analogy with the
undriven case. In the regime $t-t_w\ll t_w$
the same relation $\chi (t-t_w)=\tilde \chi (C^{st}(t-t_w))$ 
is found between the response 
and $C^{st}(t-t_w)$, as for $T>T_c$ .  
The aging part of $\chi (t,t_w)$ is negligible for $t_w\to \infty$,
as without drive, resulting in a flat $\chi (C)$ in the 
aging regime $t-t_w\gg t_w$.

\end{abstract}

Understanding non-equilibrium processes is a central topic
of modern statistical physics.
Many systems in nature are observed in a persistent non-equilibrium
regime. Among these two important categories can
be identified according to the asymptotic state towards which they 
evolve. Systems subjected to a change of thermodynamic variables,
such as temperature or pressure, belong to the first class.
In this case, an equilibrium state is approached which,
however, can be hardly reached if the ergodic time grows in the
thermodynamic limit. This is usually observed when the initial
and final states are separated by a phase transition. Typical examples
are glassy materials or phase-separating systems quenched below the
critical temperature. The off-equilibrium dynamics in this case
is characterized by the aging property of the
autocorrelation function \cite{Bouchaud97}.
A different situation, on the other hand, 
is observed when external power is injected into a system
with rate $\gamma $ by doing work on it, 
as in the case of stirred or sheared materials. 
In this case a time translational invariant (TTI) stationary regime can
be entered, depending on the thermodynamic parameters,  
which is not an equilibrium state.

A characterization of non-equilibrium states has been given
in different contexts by generalizing the fluctuation-dissipation 
theorem (FDT). 
In equilibrium the FDT relates the integrated linear response function
$\chi (t-t_w)$ to a perturbation applied at time $t_w$ and the 
autocorrelation function $C (t-t_w)$ of the unperturbed system
through $\chi (t-t_w)=(1/T)[C(0)-C (t-t_w)]$. In stationary states 
these quantities still depend on
the time difference so that $\chi (t-t_w)=\tilde \chi [C (t-t_w)]$,
where $\tilde \chi $ is a non trivial function that may give
informations on the off-equilibrium properties.
Also in the late stage
evolution of aging systems the dependence of the response function
on the two times generally enters through the autocorrelation,
$\chi (t,t_w)=\tilde \chi [C (t,t_w)]$, as originally observed  
by Cugliandolo and Kurchan \cite{Cugliandolo93} in the framework of mean field
glass models. 

In these cases the fluctuation dissipation ratio (FDR)
$X(C)=-Td\chi (C)/dC$ describes deviations from equilibrium,
where $X(C)\equiv 1$, and different physical informations 
can be extracted from it. First, $T_{eff}=TX^{-1}(C)$ 
can be interpreted as an {\it effective } temperature of the system, 
different from that of the thermal bath, as recognized in 
turbulence \cite{Hohenberg89} or glass physics \cite{Cugliandolo93,
Cugliandolo97}. 
Second, the theorem by Franz, Mezard, Parisi 
and Peliti \cite{Franz98} connects the FDR of aging systems 
to the static properties of
the equilibrium state through
$dX(C)/dC\vert _{C=q}=P(q)$, where $P(q)$ is the equilibrium 
probability distribution of the overlaps \cite{Mezard87}. 
Moreover it was argued recently~\cite{Cugliandolo97,Cugliandolo97b} 
that aging and driven systems 
can be related through the off-equilibrium generalization of the FDT
because $X (C)$
is expected to be the same in the limit of small entropy production, 
namely for $t_w\to \infty $
in aging systems or $\gamma \to 0$ in driven systems.

In this Letter we study this subject in the framework of the
large-$N$ model for systems with non conserved order 
parameter~\cite{Bray94} under plain shear flow; 
this is a solvable theory which retains
the basic features of the equilibrium and off-equilibrium kinetics
of several non-disordered systems, such as liquid crystals~\cite{Orihara86},
which are relevant for many applications.
In this context the autocorrelation and the response functions
are exactly computed and the fluctuation dissipation ratio
discussed explicitly. The model is particularly suited for
a unitary discussion of the FDT in the different non-equilibrium
situations discussed above, which correspond to dynamical states
at different temperatures. In particular, for $T$
larger than a {\it critical } temperature $T_c(\gamma )$,
a stationary state is entered, for long enough times, 
characterized by a non-trivial $X(C)$. By quenching below $T_c(\gamma)$,
instead, the complex interplay between the phase-ordering
process and the external drive gives rise to peculiar aging
properties. Nonetheless, by recognizing that a separation of
the degrees of freedom into two statistically independent contributions,
responsible for the stationary $C_\gamma^{st}(t-t_w)$ and aging
$C_\gamma ^{ag}(t/t_w)$ part of the autocorrelation, can be explicitly
performed, the interpretation of the dynamics is clarified by the analogy
with the phase-ordering kinetics of the undriven case.

The model describes the system in terms of an $N$-component vectorial
order parameter field  ${\vec \varphi({\bf r}, t)}$ through
the {\it Ginzburg-Landau} hamiltonian
\begin{equation}
{\cal H}\{\vec \varphi\} = \int d^D r
\{ \frac{\kappa}{2} \mid \nabla \vec \varphi \mid^2 \ + V(\vec \varphi ) \}
\label{eqn1}
\end{equation}
with
\be
V(\vec \varphi)= -\frac{r}{2} \mid \vec \varphi \mid ^2 + 
\frac{u}{4N} \left (\mid \vec \varphi \mid ^2 \right )^2.
\ee
where $r,u$ and $\kappa $ are positive constants.
When a convective field ${\bf v}$ is imposed
the dynamics of the generic component $\alpha $ of a non-conserved
order parameter is described by the equation
\begin{equation}
\frac {\partial \varphi _\alpha} {\partial t} + {\bf \nabla } 
\cdot (\varphi _\alpha {\bf v}) =
- \Gamma   \frac {\delta {\cal H}}{\delta \varphi _\alpha} + \eta _\alpha
\label{eqn2}
\end{equation}
where $\Gamma$ is a transport coefficient and the
gaussian stochastic variable $ \vec \eta $, representing
thermal noise, obeys 
$\langle \eta _\alpha ({\bf r}, t) \rangle =0$ and 
$\langle \eta _\alpha ({\bf r}, t) \eta _\beta ({\bf r'}, t')\rangle =
2 T \delta _{\alpha ,\beta} \Gamma \delta({\bf r} -{\bf r}') \delta(t-t')$,
with $<\dots >$ denoting ensemble averages.
In the following we will set $\Gamma =1$.
The velocity ${\bf v}$ is chosen to be
a planar Couette or shear flow ${\bf v} = \gamma y {\bf e}_x$,
where $\gamma$ is the spatially homogeneous shear rate,
and ${\bf e}_x$ is a unit vector in the flow direction.
In the large-$N$ limit summing over infinitely many components
effectively realizes an ensemble average and the substitution
\be
\frac{1}{N}\mid \vec \varphi ({\bf r},t) \mid ^2=
\frac{1}{N}\sum _{\alpha =1}^N \varphi _\alpha ^2 
\to \langle \varphi _\alpha ^2 \rangle
\ee
becomes exact~\cite{Emery75}. Given space-translational invariance the average
squared field fluctuation $S(t)=\langle \varphi _\alpha ^2 \rangle$
does not depend on the spatial coordinates and~(\ref{eqn2}) 
is formally linearized. In momentum space it reads 
\begin{equation}
\frac {\partial \varphi({\bf k},t)} {\partial t} -
\gamma k_x \frac {\partial \varphi({\bf k},t)} {\partial k_y} =
- [k^2 + u S(t) -r]\varphi({\bf k},t)+  \eta(\vec k, t)
\label{eqn6}
\end{equation}
where we have dropped the component index, due to internal symmetry,  
$\langle \eta ({\bf k}, t) \rangle  = 0 $ and
$\langle \eta({\bf k}, t) \eta({\bf k}', t')\rangle =
2 T (2 \pi)^D \delta({\bf k} +{\bf k}') \delta(t-t')$ .
The formal solution of Eq.(\ref{eqn6}) can be obtained from the knowledge
of the configuration at an arbitrary time $t_0$ by the method of
characteristics 
\be
\varphi({\bf k},t)= \varphi \left [{\bf {\cal K}}(t-t_0), t_0
\right ]
\sqrt {\frac{g(t_0)}{g(t)}}e^{- \int _0 ^{t-t_0} [{\cal K}^2(z)]dz}
+ \int _{t_0} ^t \sqrt{\frac{g(z)}{g(t)}}
e^{ - \int _0 ^{t-z} [{\cal K}^2(s)]ds }
\eta \left [{\bf {\cal K}}(t-z),z\right ] dz
\label{solution}
\ee
where ${\bf {\cal K}}(z)={\bf k}+\gamma z k_x {\bf e}_y$ and
$g(t)=\exp \left \{ 2 \int _0 ^t [uS(t')-r] dt' \right \}$.
Eq.~(\ref{solution}) allows the computation of the structure factor
${\cal C}_\gamma ({\bf k},t)=\langle \varphi ({\bf k},t)\varphi (-{\bf k},t) \rangle$
and the relation 
\be
S(t)  =  \int   \frac {d ^D k}{(2\pi)^D}
e^{-\frac{k^2}{\Lambda^2}} {\cal C}_\gamma ({\bf k},t),
\label{eqn7}
\end{equation}
where $\Lambda $ is a regularization cut-off, 
provides the self-consistent closure 
of the model. The knowledge of $g(t)$, therefore, amounts to the
solution of the theory.

The existence of {\it critical} temperature was shown in~\cite{Gonnella2000}.
For small $\gamma /\Lambda ^2$ in $D=3$ the critical temperature behaves
as $T_c(\gamma )=T_c(0)[1+a\sqrt {\gamma \Lambda ^2}]$, where $a$ is a 
positive constant. In general, it can be shown that, in this model, 
$T_c(\gamma )$ is a monotonously increasing function of the shear rate. 
We will consider quenches from an high temperature disordered state
to a final temperature $T_F$.
In the following we will briefly discuss the most relevant case $D=3$.
A complete analysis, containing also more
technical details, will appear elsewhere~\cite{Lavorolungo}.
$g(t)$ can be evaluated by Laplace transform techniques,
similarly to the case $\gamma =0$~\cite{Newman90}, yielding
\be
g(t)=  \left \{ \begin{array}{ll}
                   g_+ \, \, e^{s_0t}        
                              & \mbox{, for $T_F > T_c(\gamma)$} \\
                   g_c \, \,
                              & \mbox{, for $T_F=T_c(\gamma)$} \\
		   g_- \, \, t^{-\frac{5}{2}}     
			      & \mbox{, for $T_F<T_c(\gamma)$,}
                   \end{array}
              \right .
\label{2.12}
\ee
where $g_+ , g_c, g_- $ are constants. $s_0(T)$ is
related to the longitudinal coherence length by $s_0=2\xi (T)^{-2}$.
These behaviors are reminiscent of the case without shear~\cite{Newman90},
where the algebraic decay of $g(t)$ below the critical
temperature is a signature of the
power-laws characteristic of the phase-ordering process. 
Then, in the case of a quench from
high temperature to $T_F$ below $T_c(\gamma )$, the effects of shear 
change the phenomenology of the phase-ordering process but
do not interrupt it. Moreover, since $T_c(\gamma )>T_c(0)$,
complete phase-ordering is always observed for each $T_F<T_c(0)$,
regardless of the presence of shear. 
This is an important difference
with respect to the case often considered~\cite{Cugliandolo97b}  
where the aging process is interrupted by a power input. This is what happens
when we shake, for instance, a mixture of water and oil to make droplets
smaller. Actually the possibility of stabilizing a state with a finite
coherence length by shearing a phase-ordering system  
is a debated question~\cite{Hashimoto94,Cavagna00} which may depend
crucially on hydrodynamic effects~\cite{Onuki97}.

  
From the knowledge of $g(t)$ two-time quantities can be computed.

\vspace {3mm}
{\boldmath $T_F>T_c(\gamma )$}
\vspace {3mm}

For quenches to $T_F>T_c(\gamma )$ 
the autocorrelation function $C_\gamma(t,t_w)=
(2\pi )^{-2D}\int d^Dk d^Dk'\\
\langle \varphi ({\bf k},t)\varphi ({\bf k}',t_w)\rangle
\exp\{-(k^2+k'^2 )/(2\Lambda ^2)\}$ 
is a time translational
invariant quantity, $C_\gamma(t,t_w)=C_\gamma^{st}(t-t_w)$, 
for $t_w\gg 1/s_0$. 
From~(\ref{solution}) one finds 
\be
C_\gamma^{st}(t-t_w)=\frac {4T_F}{(8\pi )^{3/2}} \int _{\frac{t-t_w}{2}}^\infty
\frac{e^{s_0z}\left (z +\frac{1}{2\Lambda^2}\right )^{-3/2}}
{\sqrt{4 +\frac{1}{3}\gamma ^2 z^2 +
\frac{\gamma ^2}{2} (t-t_w) ^2 \left [1-\frac{(t-t_w)^2}{8z^2} \right ]}} dz.
\label{autosopra}
\ee

\begin{figure}
\includegraphics*[width=11 cm,height=8 cm]{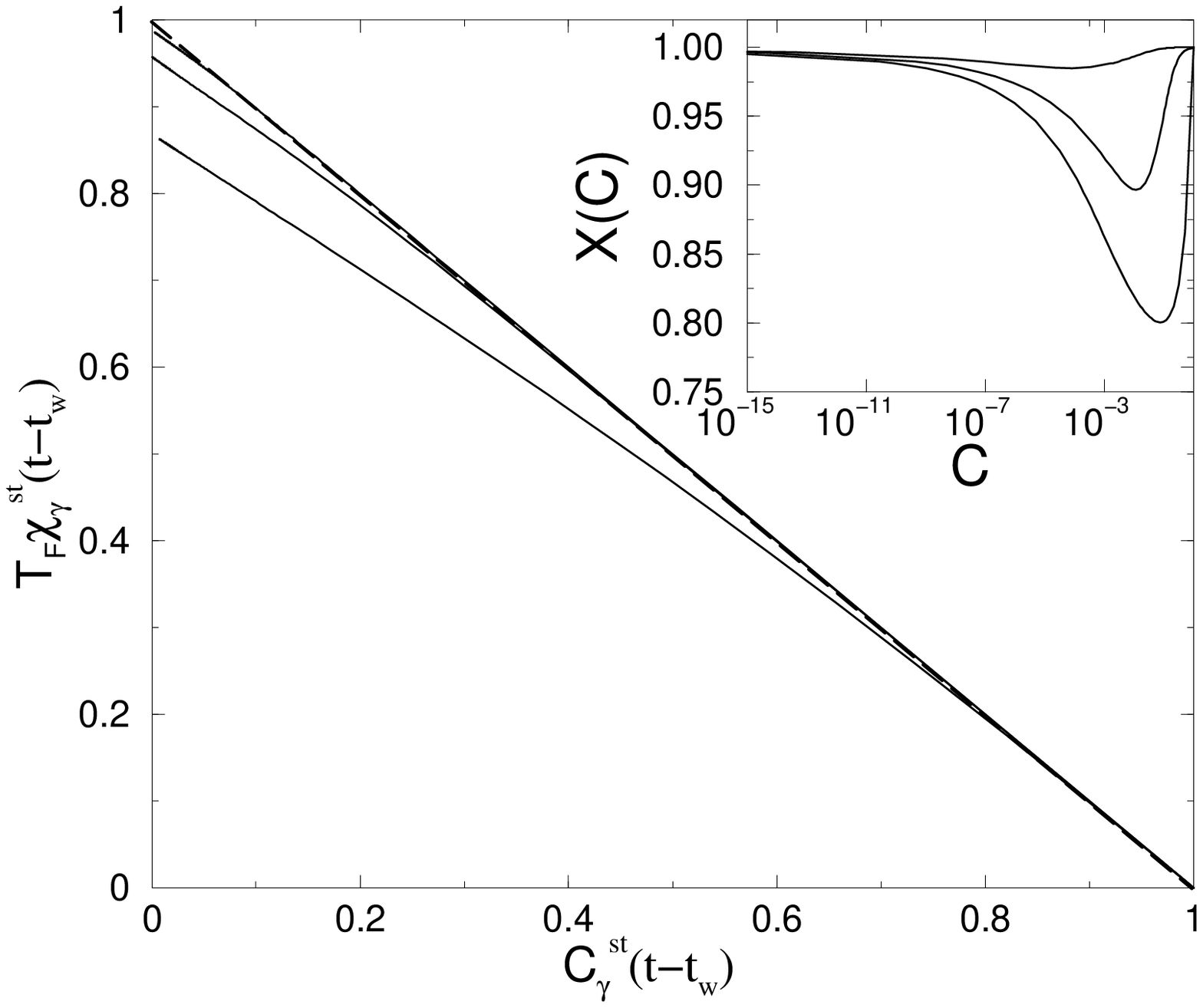}
\caption{$T_F\chi ^{st}_\gamma (t-t_w)$ is plotted against 
$C^{st}_\gamma (t-t_w)$ for a quench to a final temperature
$T_F=(21/20)T_c(\gamma )$, with $t_w=10^5$ and $\Lambda=1$. 
Solid lines correspond to 
$\gamma =1, 10^{-1}, 10^{-2}$, bottom up. The dashed line, which is
almost indistinguishable from the case $\gamma =10^{-2}$,
is the equilibrium behavior $\gamma =0$.
The inset shows $X(C)$.}
\label{fdtsopra}
\end{figure}

The next point is the calculation of the linear response function
${\cal R}({\bf k},t;{\bf k}',t') =$ \\ $(2 \pi)^D
\delta \langle \varphi({\bf k}, t)/
\delta h(-{\bf k}', t')\mid _{h({\bf k}, t)= 0 }$ (with $t\geq t'$)
to a perturbation $\vec h ({\bf r},t)$ entering the hamiltonian through
$-\int d^D r \vec h ({\bf r},t) \cdot \vec \varphi ({\bf r},t)$.
From this quantity the integrated response function is obtained
as $\chi_\gamma (t,t_w)=\int _{t_w} ^t dt' (2\pi )^{-2D} \int d^Dk d^D k'
{\cal R}({\bf k},t;{\bf k}',t')\exp\{-(k^2+k'^2 )/(2\Lambda ^2)\}$. 
For $t_w\gg 1/s_0$ it can be
shown that $\chi_\gamma (t,t_w)=\chi_\gamma ^{st}(t-t_w)$ with
\be 
\chi_\gamma ^{st} (t-t_w) = \frac { \Gamma(\frac{3}{2})}{2\pi^{2}}
 \int _0 ^{t-t_w}  dz
e^{-\frac{s_0z}{2}}\frac {\left (z+ \frac{1}{\Lambda^2 }\right )^{-3/2}}
{\sqrt{4 + \frac{1}{3} \gamma^2 z^2}}.
\label{risposta}
\ee
Writing the argument of $\chi_\gamma (t-t_w)$ 
in terms of $C_\gamma^{st}(t-t_w)$ through~(\ref{autosopra})
one obtains the fluctuation dissipation relation $\chi_\gamma (t-t_w)= \tilde
\chi_\gamma (C_\gamma^{st})$ plotted in Fig.~\ref{fdtsopra}. 
The comparison with
the case without shear, regulated by equilibrium FDT, shows that for
small $\gamma (t-t_w)$ the two curves collapse but, for larger 
time differences the lines with $\gamma \neq 0$ bend and 
lower asymptotic values are reached. The behavior of the FDR
is shown in the inset of Fig.~\ref{fdtsopra}. 
Starting from $X (0)=1$, this quantity decreases to a minimum and then 
raises towards $X(C)=1$ again when $C\to 1$. Therefore one finds
a different and more complex behavior with respect to the 
two-temperature scenario 
proposed for driven supercooled fluids~\cite{Berthier2001}. 

\vspace {3mm}
{\boldmath $T_F=T_c(\gamma )$}
\vspace {3mm}

For quenches at the critical temperature the asymptotic behavior
of the two time quantities considered above is the same as for 
$T_F>T_c (\gamma )$,
with $s_0=0$ in~(\ref{autosopra}) and~(\ref{risposta}).
In particular $C_\gamma(t,t_w)$ and $\chi_\gamma (t,t_w)$ in the large 
$\gamma t_w$ limit depend only on $t-t_w$.

\vspace {3mm}
{\boldmath $T_F<T_c (\gamma )$}
\vspace {3mm}

The situation is more complex when quenches below $T_c(\gamma )$
are considered.
In this case, using~(\ref{solution}), 
in the large $t_w$ limit the autocorrelation function 
can be separated into two
contributions, namely $C_\gamma (t,t_w)=C^{st}_\gamma (t-t_w)
+C^{ag}_\gamma (t/t_w)$,
which depend, respectively, on the difference and on the
ratio between the two times, as usually found in aging systems. 
This is the signature of a fundamental property of
the order parameter field $\vec \varphi({\bf r},t)$, which 
decomposes below $T_c(\gamma )$ into the sum of two 
independent stochastic fields 
$\vec \varphi ({\bf r},t)=\vec \psi ({\bf r},t)+\vec \sigma ({\bf r},t)$
responsible for the stationary and
aging properties of the kinetics, as discussed in~\cite{Mazenko88}.
In systems with  
a scalar order parameter~\cite{Corberi2001} the physical
interpretation of this decomposition goes as follows:
given a configuration at time $t_0$ well inside the late stage scaling
regime, one can distinguish degrees of freedom $\psi ({\bf r},t)$ 
in the bulk of domains from those $\sigma ({\bf r},t)$ pertaining 
to the interfaces. The first ones are driven by thermal fluctuations
and behave locally as in equilibrium; the second ones 
retain memory of the noisy initial condition, and produce the aging behavior.
In the limit $N\to \infty$ the separation of $\varphi ({\bf k},t)$
can be explicitly constructed,
identifying in~(\ref{solution}) the two contributions
\be
\psi({\bf k},t)=\lim _{t_0\to \infty} 
\int _{t_0} ^t \sqrt{\frac{g(z)}{g(t)}}
e^{ - \int _0 ^{t-z} [{\cal K}^2(s)]ds }
\eta \left [{\bf {\cal K}}(t-z),z\right ] dz
\label{psi}
\ee
and
\be
\sigma({\bf k},t)=\lim _{t_0\to \infty} \varphi \left [{\bf {\cal K}}(t-t_0),
t_0 \right ]
\sqrt {\frac{g(t_0)}{g(t)}}e^{-\int _0 ^{t-t_0} [{\cal K}^2(z)]dz} 
\label{sigma}
\ee
in complete analogy with the case $\gamma =0$~\cite{Corberi2002}. 
For large-$N$, where topological defects are unstable and
domains are not well defined, the recognition of the degrees of freedom 
associated to $\psi({\bf k},t)$ and $\sigma({\bf k},t)$ is not
straightforward as for scalar systems. 
Nevertheless, the possibility of the
splitting~(\ref{psi},\ref{sigma}) of the order parameter 
indicates that the same fundamental
property is shared by systems with different $N$.

Building the autocorrelation functions of the new fields
$C_\gamma ^\psi (t,t_w)=(2\pi )^{-2D}\int d^Dk d^Dk' \\
\langle \psi ({\bf k},t)\psi ({\bf k}',t_w)\rangle
\exp\{-(k^2+k'^2 )/(2\Lambda ^2)\}$ and 
$C_\gamma ^\sigma (t,t_w)=(2\pi )^{-2D}\int d^Dk d^Dk'
\langle \sigma ({\bf k},t)\sigma ({\bf k}',t_w)\rangle\\
\exp\{-(k^2+k'^2 )/(2\Lambda ^2)\}$ 
it is possible to prove that, for $t_w/t_0$ sufficiently large,
they coincide, respectively, with $C_\gamma ^{st}(t,t_w)$ and 
$C_\gamma ^{ag}(t,t_w)$. 

Coming back to the two-time observables
the TTI part $C_\gamma^{st}(t-t_w)$ takes the same form as at
$T_F=T_c(\gamma )$, namely~(\ref{autosopra}) 
with $s_0=0$. This quantity decays from
$C_\gamma^{st}(0)=M_\gamma^2(0)-M_\gamma^2(T_F)$ 
to zero for time differences of order $t_w$.
Here $M_\gamma(T_F) = M_\gamma(0) [1-T_F/T_c(\gamma)]^{1/2}$
is the spontaneous magnetization at stationarity.
The aging part, instead, is given by
\be
C_\gamma ^{ag}(\frac {t}{t_w})= 2\sqrt{\frac {2}{3}}\,M_\gamma^2(T_F)
\left (1+\frac{t}{t_w}\right )^{-\frac{3}{2}}
\frac {1}{\sqrt{ \frac{4}{3} \frac {2-\left (1-\frac{t_w}{t}\right )^3}
{1+\frac{t_w}{t}}-
\frac {\left [2-\left (1-\frac{t_w}{t}\right )^2\right ]^2}
{\left (1+\frac{t_w}{t}\right )^2} }},
\label{autosigma}
\ee
showing that $C_\gamma ^{ag}$ stays approximatively
constant, $C_\gamma ^{ag}(t/t_w)\simeq M_\gamma^2(T_F)$,
for $t-t_w\ll t_w$ and decays 
for large time differences as $(t_w/t)^\lambda$,
with $\lambda =5/4$, as opposed to the undriven case
where $\lambda =3/4$.
Computing the response function it is found that 
the aging part is negligible for large $t_w$ 
and $\chi (t,t_w)$ behaves as at $T_c(\gamma )$, 
so that~(\ref{risposta}) with $s_0=0$ is obeyed.

\begin{figure}
\includegraphics*[width=11 cm,height=8 cm]{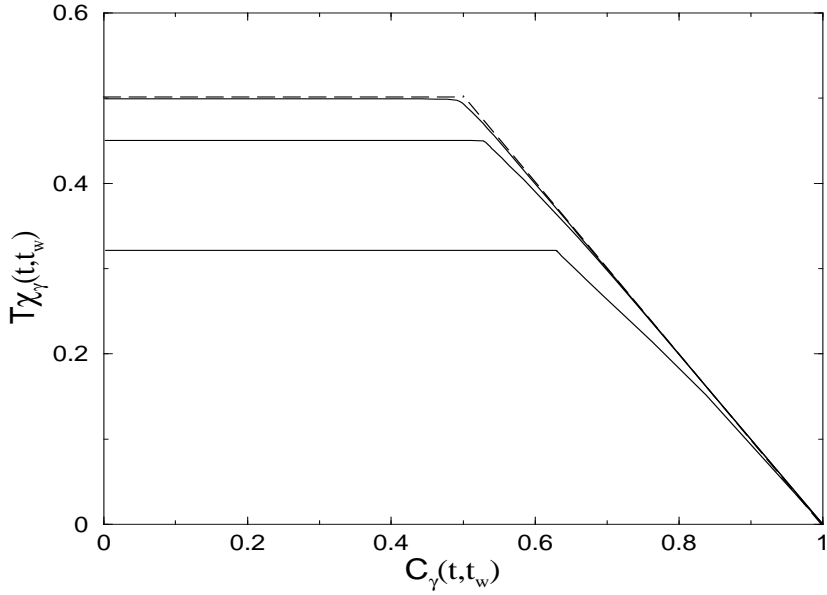}
\caption{$T_F\chi _\gamma (t,t_w)$ is plotted against 
$C_\gamma (t,t_w)$ for a quench to a final temperature
$T_F=T_c(\gamma )/2$, with $t_w=10^5$ and $\Lambda=1$. 
Solid lines correspond to 
$\gamma =1, 10^{-1}, 10^{-2}$, bottom up. The dashed line 
is the case with $\gamma =0$.}
\label{fdtsotto}
\end{figure}

We turn now to discuss the relation between 
$\chi _\gamma (t,t_w)$ and $C_\gamma (t,t_w)$. 
In the case without shear~\cite{Corberi2002}, the equilibrium FDT relates
$\chi^{st}(t-t_w)$ and $C^{st}(t-t_w)$ by 
$T_F\chi^{st}(t-t_w)=C^{st}(0)-C^{st}(t-t_w)$ at every 
temperature.
In the range of times 
$t-t_w \ll t_w$ (stationary regime), 
when $C^{st}(t-t_w)$ decays to zero, 
$C^{ag}(t/t_w)\simeq M^2(T_F)$ and 
$T_F\chi^{st} (t,t_w)$ 
grows to its saturation
value $M^2(0)-M^2(T_F)$.
Then, plotting $T_F\chi (t,t_w)$ against $C(t,t_w)$ one gets 
the straight line with negative slope on the right of Fig.~\ref{fdtsotto}. 
For larger time differences,
in the aging regime $t-t_w\gg t_w$, $\chi (t,t_w)$ is a constant, 
because the aging part of the response is negligible
for $t_w\to \infty $. In the meanwhile
$C^{ag}(t/t_w)$ slowly decreases, and
the characteristic horizontal line of phase ordering is found.
When the drive is switched on, $\chi_\gamma ^{st}(t-t_w)$ and 
$C_\gamma^{st}(t-t_w)$
are not related by equilibrium FDT but, instead, through 
$\tilde \chi_\gamma (C_\gamma^{st})$ (Fig.~\ref{fdtsopra}). 
Then in the parametric
plot of $\chi_\gamma (t,t_w)$ vs $C_\gamma(t,t_w)$ (Fig.~\ref{fdtsotto}) 
one finds,
in place of the straight line with negative slope, the curve 
$\tilde \chi_\gamma (C_\gamma^{st})$ which corresponds 
to the simple translation
of the one plotted in Fig.~\ref{fdtsopra}. 
For $C_\gamma(t,t_w)<M_\gamma^2(T_F)$
the plot is flat for the same reasons as for $\gamma =0$. 
However, due to the bending
of $\tilde \chi_\gamma (C_\gamma^{st})$, the asymptotic value 
$\chi_\gamma (\infty)$
is different from the case $\gamma =0$. It is interesting to observe
that the relation $T_F\chi (\infty) =M^2(0)-M^2(T_F)$ characteristic
of the undriven case is generalized to 
$T_F\chi_{\gamma/2} (\infty) =M_{\gamma}^2(0)-M_{\gamma}^2(T_F)$.
Therefore a relation analogous to the case $\gamma =0$ holds between
the response function of a system with shear rate $\gamma /2 $ and
the stationary magnetization in the presence of a shear rate $\gamma $
twice as large.
This parameter-free analytical prediction is amenable of experimental
testing. 

In summary, in this Letter we have studied the effects of a shear flow on 
the fluctuation-dissipation relation
in the large-$N$ model for phase-ordering.
This approach allows for the analytical computation
of two times quantities and the explicit identification of the 
stochastic fields $\psi ({\bf r },t)$, $\sigma({\bf r },t)$ 
responsible for the stationary and aging properties of the
kinetics below $T_c(\gamma )$.  
The FDT relating the stationary parts
of the correlation and response function in the undriven case
is generalized under shear into a non-trivial relation
$\chi ^{st}_\gamma (t-t_w)=\tilde \chi _\gamma (C^{st}(t-t_w))$,
which is obeyed at every temperature. Below $T_c(\gamma )$
the lack of time translational invariance is testified by the
aging contribution $C^{ag}(t/t_w)$ to the correlation function,
which behaves asymptotically as $(t_w/t)^\lambda$, where
$\lambda =(D+2)/4$ is a generalization of the 
Fisher-Huse exponent~\cite{Fisher1988} to sheared systems. The aging
part of the response, on the other hand, is
always negligible in the large $t_w$ limit, resulting in a
flat part of the fluctuation dissipation plot.
The global picture provided by the model is closely related to
the one observed without shear, the most notable modification being
the different relation between $\chi ^{st}_\gamma (t-t_w)$
and $C^{st}_\gamma (t-t_w)$. The comparison with the case $\gamma =0$ 
will be discussed further in~\cite{Lavorolungo}.

The applicability
of the two-temperature scenario proposed in~\cite{Berthier2001} for
supercooled fluids to different slowly relaxing systems and,
in particular, to the phase-ordering kinetics is a debated~\cite{Onuki97} 
issue which is worth of further studies. 
According to this scenario an external input of energy stabilizes
an aging system into a power-dependent stationary state. In the
small $\gamma $ limit the fluctuation dissipation plot of this
state is a broken line. This provides the identification of two well 
defined temperatures: The bath temperature $T_F$ and an {\it effective }
temperature of the system.
The results of this Letter allows this problem to be addressed in  
the context of the large-$N$ model.  We find that
a phase-ordering system is not stabilized by shear.
This result agrees with Ref.~\cite{Cavagna00},
where a different approximation scheme
is developed.
A similar behavior was found in~\cite{Crisanti87} for
the stability of the ferromagnetic phase of the asymmetric spin-glass
model in the mean-field limit.
This feature is at variance with the basic 
assumption of the scenario proposed for supercooled fluids.
Furthermore, the behavior of $X(C)$ shown in the inset of Fig.~\ref{fdtsopra}
shows that a two-temperature
interpretation of the fluctuation dissipation relation
does not hold in this case.

The relevance of these results to real physical systems where $N$
is finite and hydrodynamic effects may play a relevant role
is a crucial point that deserves further investigations. 

\stars

We acknowledge support by the TMR network contract ERBFMRXCT980183,
MURST(PRIN 2000) and INFM PRA-HOP 1999.

\dag corberi@na.infn.it ; *gonnella@ba.infn.it

\ddag lippiello@sa.infn.it ; \S zannetti@na.infn.it

\end{document}